\documentclass[11pt, oneside]{article}

\usepackage{amssymb}
\usepackage{amsmath}
\usepackage{amsthm}
\usepackage{amstext}
\usepackage{amscd}
\usepackage{enumerate}

\newcommand*{\cH}{\mathcal{H}}
\newcommand*{\ket}[1]{| #1 \rangle}
\newcommand*{\bra}[1]{\langle #1 |}
\newcommand*{\spr}[2]{\langle #1 | #2 \rangle}

\newcommand*{\stwo}{\frac{1}{\sqrt{2}}}

\newcommand*{\cG}{\mathcal{G}}
\newcommand*{\cbG}{\mathcal{G}^g}
\newcommand*{\csG}{\mathcal{G}^s}
\newcommand*{\ve}[1]{\ket{e_{#1}}}
\newcommand*{\veb}[1]{\bra{e_{#1}}}
\newcommand*{\vf}[1]{\ket{f_{#1}}}
\newcommand*{\vg}[1]{\ket{g_{#1}}}
\newcommand*{\borg}[1]{b^{(#1)}}
\newcommand*{\bone}[1]{\bar{b}^{(#1)}}
\newcommand*{\bmax}[1]{\tilde{b}^{(#1)}}
\newcommand*{\bfin}[1]{\hat{b}^{(#1)}}

\newcommand*{\bb}{\bar{b}}

\newcommand*{\bN}[1]{\{1,\ldots, #1\}}
\newcommand*{\Bcl}{B^{\mathrm{cl}}}
\newcommand*{\Bqm}{B^{\mathrm{qm}}}
\newcommand*{\GF}{\mathrm{GF}}
\newcommand*{\kgs}{\ket{\Psi}}
\newcommand*{\bs}{\Phi}
\newcommand*{\kbs}{\ket{\bs}}
\newcommand*{\taucl}{\tau_{\mathrm{cl}}}
\newcommand*{\tauqm}{\tau_{\mathrm{qm}}}

\theoremstyle{plain}
\newtheorem{theorem}{Theorem}
\newtheorem{lemma}[theorem]{Lemma}

\theoremstyle{definition}
\newtheorem{definition}[theorem]{Definition}

\title{\bf Towards Characterizing the Non-Locality of Entangled Quantum
  States}
 
\date{October, 2002}

\author{
  Renato Renner\thanks{
    Computer Science Department, ETH Z\"urich, 
    CH-8092 Z\"urich, Switzerland. 
    E-mail: {\tt renner@inf.ethz.ch}.}  
  \ and Stefan Wolf\thanks{
    D\'epartement d'informatique et recherche op\'erationelle, 
    Universit\'e de Montr\'eal, 
    6128 succ Centre-Ville, Montr\'eal, Qu\'ebec, H3C 3J7, Canada.
    E-mail: {\tt wolf@iro.umontreal.ca}.}}

\begin{document}

\maketitle

\begin{abstract} 

  \noindent The behavior of entangled quantum systems can generally not be
  explained as being determined by shared classical randomness.  In
  the first part of this paper, we propose a simple game for $n$
  players demonstrating this non-local property of quantum mechanics:
  While, on the one hand, it is immediately clear that classical
  players will lose the game with substantial probability, it can, on
  the other hand, always be won by players sharing an entangled
  quantum state. The simplicity of the classical analysis of our game
  contrasts the often quite involved analysis of previously proposed
  examples of this type.
  
  In the second part, aiming at a quantitative characterization of the
  non-locality of $n$-partite quantum states, we consider a general
  class of $n$-player games, where the amount of communication between
  certain (randomly chosen) groups of players is measured.  Comparing
  the classical communication needed for both classical players and
  quantum players (initially sharing a given quantum state) to win
  such a game, a new type of separation results is obtained.  In
  particular, we show that in order to simulate two separated qubits
  of an $n$-partite GHZ state at least $\Omega(\log \log n)$ bits of
  information are required.

\end{abstract}

\section{Introduction}

\subsection{Quantum Entanglement vs. Classical Correlation}

Consider an entangled quantum state shared between $n$~parties, each
belonging to a separate, dynamically isolated system.\footnote{One
  often considers systems which are spatially separated (after the
  common quantum state has been prepared), such that, according to the
  theory of relativity, there is no causal connection between any two
  events of interest belonging to different systems.} It is a known
fact of quantum mechanics that entanglement cannot be used to achieve
communication, i.e., no information can be exchanged between these
parties.  Nevertheless, according to Bell's well known
theorem~\cite{B64}, the outcomes of local measurements on the $n$
systems are generally correlated in a non-classical way. This means
that the (classical) measurement outcomes can \emph{not} be simulated
by parties only sharing classical information instead of quantum
entanglement. Consequently, albeit not allowing for communication, a
shared quantum state might help the $n$~parties to accomplish certain
tasks.

Aiming at understanding the nature of quantum entanglement, it is
instructive to study simple examples of such tasks.  To this end, we
will consider \emph{games} between $n$ collaborating \emph{players}.
A game is \emph{won} if the players' answers satisfy a given condition
(possibly depending on a query).  To generate their outputs, the
players are allowed to perform arbitrary local (quantum) computations
(in particular, they are computationally unbounded) but the
communication between them is subject to restrictions.

It turns out that there are games which can always be won by players
sharing an entangled quantum state while any classical strategy to win
fails with some positive probability. These are often called
\emph{pseudo-telepathy games} since the behavior of successful players
cannot be explained classically without assuming some hidden extra
communication between them. A nice example for such a game has been
presented in~\cite{BCT99} where two collaborating players must answer
a query in a somehow correlated way without being allowed to
communicate.

In Section~\ref{sec:simpgame}, we propose another particularly simple
game of this type involving $n \geq 5$ collaborating players: Two
randomly picked players receive a bit $b$ (being chosen by the
remaining $n-2$ players) which they can either flip or leave
unchanged. The game is won if the two players, without being allowed
to communicate (in particular, none of them knows who the other one
is), behave differently, meaning that \emph{exactly} one of them flips
the bit $b$.  Obviously, there is no classical strategy for winning
the game with certainty. Indeed, a simple argument (see
Section~\ref{sec:simpcl}) shows that the probability to lose is
substantial (roughly $1/4$ for $n$ large). This stands in contrast to
other similar games, where the classical analysis is often quite
involved or where only asymptotic results are proven.\footnote{E.g.,
  for the mentioned two-player game from~\cite{BCT99}, the probability
  to lose has been shown (based on graph-theoretical results) to be
  positive for any classical strategy, but no lower bound for this
  probability is known yet (cf.~\cite{GW02,GTW02}).} On the other
hand, our game can be won with certainty by players initially sharing
a GHZ state (Section~\ref{sec:simpqm}).

\subsection{Non-Local Information} \label{sec:intrnl}

An $n$-partite quantum state $\kgs$ can be seen as a resource
consisting of $n$ components, each of them taking a classical input
(the measurement basis) and generating a classical output (the outcome
of the measurement performed on the respective part of $\kgs$).  As
described above, such a resource is generally more powerful than its
classical counterpart, i.e., $n$ separated components sharing purely
classical information. This non-classical property of quantum states
is often called \emph{non-locality} or \emph{non-local information of
  $\kgs$}.


The non-local information of an $n$-partite quantum state $\kgs$ over
$n$~subsystems can be characterized by the minimal amount of
communication between $n$ separated classical systems needed for
simulating the behavior of the respective quantum subsystems.  While,
in general, it is not clear how to determine this communication, it
turns out that the games mentioned in the previous section are useful
to find certain bounds: Consider a game which can always be won by
players sharing the quantum state $\kgs$.  The non-local information
of $\kgs$ is then lower bounded by the minimal \emph{additional}
communication being necessary for classical players to win this game.

To obtain a \emph{real-valued measure} for the non-local information
of a state $\kgs$, the communication needed for its simulation has to
be quantified. This can be done in several ways, but any concrete
measure merely unveils certain aspects of this communication, and,
consequently, does not fully characterize the non-local information
contained in $\kgs$. It thus seems that, in order to understand the
nature of non-locality, it is worth considering different types of
such measures.

One possibility is to rely on the definition of communication
complexity introduced by Yao~\cite{Y79}. In this setting, the
communication between \emph{two} parties is simply characterized by
the number of bits exchanged between them. In a generalization to
$n>2$ parties, any message sent by a player is considered as being
broadcasted, i.e., a bit sent to all players only counts once. Cleve
and Buhrman~\cite{CB97} were the first to propose an analysis of
entanglement based on this communication model. It could be
shown~\cite{BCD01} that the \emph{communication complexity of
  functions} (which can be seen as the communication necessary to win
certain games\footnote{Let $f$ be a function of $n$~variables.  The
  communication complexity of $f$ is defined as the minimal amount of
  communication necessary for $n$~players, each holding one input
  variable $x_i$, to compute the value of $f(x_1, \ldots, x_n)$. The
  setting thus corresponds to a game which is won if each player
  outputs the correct value of $f(x_1, \ldots, x_n)$.}) is generally
larger for purely classical players than for players sharing entangled
quantum states.  In a variety of papers, e.g.,
\cite{ASTVW98,BCW98,BDHT99,R99,BCD01} (see~\cite{BCD01} or~\cite{B01}
for a survey), the communication complexity for both the quantum and
the classical case, and in particular the gap between them, has been
studied extensively.

In this paper, we derive a slightly different type of separation
results. The idea is to not only consider the overall entanglement of
a quantum state, but also the non-local information contained in
certain of its parts. For instance, given a state $\kgs$ defined on
$n$~subsystems, one might be interested in the non-local correlation
between any two of the subsystems and, additionally, the dependence of
this correlation from the information contained in the other $n-2$
subsystems.\footnote{The same questions arise in classical information
  theory: Given $n$ random variables $X_1, \ldots, X_n$, one usually
  is not only interested in the overall correlation between them, but
  also in the correlation between two random variables (the mutual
  information $I(X_i; X_j)$), possibly conditioned on a third one (the
  conditional mutual information $I(X_i; X_j|X_k)$).}

Our results are, similar to the mentioned results based on Yao's
model, derived from an analysis of the classical communication
necessary to win certain games. This communication is however
quantified in a different way: Instead of considering all messages
exchanged between the players, only bits transmitted between certain
groups of players are counted (ignoring all communication within these
groups).


This concept is introduced more formally in Section~\ref{sec:quant}.
As an example, we show that, in order to win the $n$-player game from
Section~\ref{sec:simpgame} classically, the amount of information that
the two chosen players must receive is at least $\Omega(\log \log n)$
bits (Section~\ref{sec:firstsep}). This is in contrast to the quantum
case where one bit always suffices, given that the $n$~players share a
GHZ state.  Consequently, to simulate local measurements on each of
two arbitrary qubits of an $n$-partite GHZ state (where the
measurement bases might depend on the other $n-2$ qubits),
$\Omega(\log \log n)$ bits of additional information are needed. Note
that this amount can be arbitrarily large (for large $n$), while the
two simulated systems are both two-dimensional.

To obtain even stronger separation results, we propose a generalized
version of the game from Section~\ref{sec:simpgame}. It is shown
(Section~\ref{sec:gengame}) that the number of bits which have to be
exchanged between (certain groups of) $n$ classical players in order
to win this generalized game is at least $\Omega(\log n)$. On the
other hand, if the players are allowed to share a GHZ state, one
classical bit of communication still suffices.

\section{A Simple Pseudo-Telepathy Game} \label{sec:simpgame}

\subsection{The Game and Its Rules} 

Consider the following game $\csG_n$ involving $n\geq 5$ collaborating
players $P_1,\ldots,P_n$.  First, two of the players are chosen
randomly in such a way that neither of them knows who the other one
is. (The non-chosen, remaining, $n-2$ players can be allowed to know
which pair of players was chosen.) In the following, we will (without
loss of generality) call the two chosen players~$P_1$ and~$P_2$.

The remaining players are now allowed to communicate and generate one
``hint'' bit $b$, which they say out loud (in particular, $P_1$ and
$P_2$ can hear the bit). The chosen players~$P_1$ and~$P_2$ must then
independently (i.e., no communication between them is allowed)
generate a bit $b_1$ and $b_2$, respectively. The game is won simply
if $b_1 \neq b_2$.

We will show that this game can be won with probability at most
(roughly) $75\%$ classically (if $n$ is large enough), but with
probability~$1$ (for any value of $n$) if the players can share
quantum information.

\subsection{Classical Analysis} \label{sec:simpcl}

Let us first consider a classical setting where the players rely on
arbitrary {\em classical\/} (but no quantum) information which might
have been shared during an initialization phase (before the start of
the game).

Each player has a fixed strategy defining his behavior for the case he
is chosen. While this strategy might in general be probabilistic,
i.e., depend on some randomness, we can, without loss of generality,
assume that this randomness is fixed before the game starts. This
means that by the time the player is chosen, his strategy is
deterministic.

Once a player is chosen, the only information he gets is the hint bit
$b$. For any given (deterministic) strategy, this bit $b$ thus
completely determines his output. Obviously, there exist exactly {\em
  four\/} possible strategies, namely to output $0$, $1$, $b$, or
$\overline{b}$ (where $\overline{b}$ denotes the complement of $b$).

If the strategies of the two chosen players are the same, they will
clearly output the same bit and the game is lost. (Otherwise, if their
strategies are different and if the remaining players know these
strategies, they can always win.) Finding the minimal probability of
losing the game thus amounts to determining the minimal probability of
the event that two players with the \emph{same} strategy (where four
strategies are possible) are picked.

For $n = 4k+r$ players (where $k$, $r$ are integers, $1 \leq k$, and
$0\leq r<4$), the probability of this event is at least
\[
    p(n)
  = (4-r)\cdot \frac{k}{n}\cdot \frac{k-1}{n-1}
    + r\cdot \frac{k+1}{n}\cdot \frac{k}{n-1}\ .
\]
We have for instance $p(5)=1/10$, $p(8)=1/7$, and $p(n)\rightarrow 1/4$ 
for $n\rightarrow \infty$.
 
\subsection{A Winning Strategy for Quantum Players} \label{sec:simpqm}

We will now show that the game can be won {\em with certainty\/} if
the players can not only share classical information, but are
additionally allowed to store a quantum state which is generated and
shared before the game starts. (During the game, the players are only
allowed to process the quantum information locally, i.e., an external
observer would not be able to detect that the players follow a quantum
strategy.)

Assume that each player~$P_i$ (for $i = 1, \ldots, n$) controls a
two-dimensional subspace $\cH_i$ of a quantum system $\cH = \cH_1
\otimes \cdots \otimes \cH_n$. Let $\{\ve{0}, \ve{1}\}$ be an
orthonormal basis of $\cH_i$ (for $i = 1, \ldots, n$). The diagonal
and the circular basis of $\cH_i$ are then given by the vectors
\begin{align}
  \vf{0} &:= \stwo (\ve{0} + \ve{1}) & 
  \vf{1} &:= \stwo (\ve{0} - \ve{1}) \\
\intertext{and}
  \vg{0} &:= \stwo (\ve{0} + i \, \ve{1}) &
  \vg{1} &:= \stwo (\ve{0} - i \, \ve{1})\ ,
\end{align}
respectively.

The quantum strategy to win the game is the following: The players
start with a so called GHZ state (see \cite{GHZ89,GHSZ90})
\begin{equation} \label{eq:ghzdef}
    \kbs
  := \frac{1}{\sqrt{2}}(
    \underbrace{
      \ve{0} \otimes \cdots \otimes \ve{0}}
        _{\text{$n$ times}}
    +
    \underbrace{
      \ve{1} \otimes \cdots \otimes \ve{1}}
        _{\text{$n$ times}}) \in \cH_1 \otimes \cdots \otimes \cH_n
\end{equation}
(being prepared before the start of the game). During the game, after
the two players have been randomly chosen, the remaining players first
measure their subsystems with respect to the diagonal basis $\{\vf{0},
\vf{1}\}$ and determine the number $k$ of players having the
measurement outcome $\vf{1}$. The parity of $k$ is then announced to
the chosen players as hint bit $b$, i.e., $b \equiv k \pmod{2}$.

Depending on the bit $b$, each of the chosen players measures his
subsystem in either the diagonal basis $\{\vf{0}, \vf{1}\}$ (if $b=1$)
or the circular basis $\{\vg{0}, \vg{1}\}$ (if $b=0$). His output is
then a bit indicating his measurement result (e.g., $0$ for $\vf{0}$
or $\vg{0}$, and $1$ otherwise).

In order to prove that, following this strategy, the players always
win the game, it suffices to verify that the measurement outcomes of
the chosen players (let them again be called~$P_1$ and~$P_2$) are
always different. Using the diagonal basis for the subsystems $\cH_3,
\ldots, \cH_n$, the players' initial state $\kbs$ can be written as
\begin{align*}
  \kbs = \quad 2^{-\frac{n-1}{2}} 
               \bigl( \ve{0} \otimes \ve{0} & \otimes
               (\vf{0} + \vf{1}) \otimes \cdots \otimes (\vf{0} + \vf{1}) \\
               + \underbrace{\ve{1} \otimes \ve{1}}_
                 {\in \cH_1 \otimes \cH_2} & \otimes
               \underbrace{(\vf{0} - \vf{1}) \otimes \cdots \otimes
               (\vf{0} - \vf{1}) \bigr)}_
                 {\in \cH_3 \otimes \cdots \otimes \cH_n}\ .
\end{align*}
Thus, obviously, the measurements performed by the remaining players
$P_3, \ldots, P_n$, getting outcomes $\vf{m_3}, \ldots, \vf{m_n}$,
respectively, project the state $\kbs$ to
\begin{equation*}
    \ket{\Phi_{m_3 \cdots m_n}} 
  = \stwo\bigl(\ve{0} \otimes \ve{0} 
    + (-1)^{\sum_{i=3}^n m_i} \ve{1} \otimes \ve{1}\bigr) \otimes 
   \ket{f_{m_3}} \otimes \cdots \otimes \ket{f_{m_n}}\ .
\end{equation*}
Note that the exponent $\sum_{i=3}^n m_i$ can be replaced by the hint
bit $b$. We are thus in one of the following situations:

\begin{enumerate}[(a)]
  \item The subsystem $\cH_1 \otimes \cH_2$ of~$P_1$ and~$P_2$ is in the
  state
    \[
      \ket{\phi^+} = \stwo (\ve{0} \otimes \ve{0} + \ve{1} \otimes
      \ve{1}) \in \cH_1 \otimes \cH_2
    \]
    and $b=0$.
    
  \item The subsystem $\cH_1 \otimes \cH_2$ is in the state
    \[
      \ket{\phi^-} = \stwo (\ve{0} \otimes \ve{0} - \ve{1} \otimes \ve{1})
      \in \cH_1 \otimes \cH_2
    \]
    and $b=1$.
\end{enumerate}

Rewriting these states in terms of the measurement bases of~$P_1$
and~$P_2$ (which according to the described strategy depend on~$b$) we
get
\begin{eqnarray*}
  \ket{\phi^+} & = & \stwo (\ket{g_0} \otimes \ket{g_1} 
                    + \ket{g_1} \otimes \ket{g_0}) \\
  \ket{\phi^-} & = & \stwo (\vf{0} \otimes \vf{1}
                   + \vf{1} \otimes \vf{0})\ .
\end{eqnarray*}
Consequently, in both cases, the measurement outcomes of~$P_1$
and~$P_2$ are always different, which concludes the proof.

\section{Quantifying Non-Local Information} \label{sec:quant}

As described in Section~\ref{sec:intrnl}, in order to characterize the
non-local information of an $n$-partite quantum state $\kgs$ in terms
of communication complexity, a measure to quantify the communication
between $n$ systems is required. We will introduce a notion of
communication complexity with respect to certain partitionings of the
$n$ systems into groups, where only the communication between these
groups is counted.

This is formalized in terms of $n$-player games (where the $n$ players
correspond to the $n$ systems). A game is a specification of both the
partitioning of the players into groups and a task which has to be
accomplished by the players (Section~\ref{sec:model}).  The group
broadcast complexity of a game is then defined as the minimal amount
of classical inter-group communication needed for the players to win
the game, i.e., to accomplish a certain task (Section~\ref{sec:bc}).
The comparison of the group broadcast complexity for both classical
and quantum players will finally lead to a new type of separation
results (Section~\ref{sec:firstsep} and~\ref{sec:gengame}).

\subsection{Games and Players} \label{sec:model}

A \emph{game $\cG_n$ for $n$ players} is defined by a probability
distribution over triples $(\sigma, q, W)$ where $\sigma$ describes a
\emph{partitioning of the players} into groups, $q$ a \emph{query} to
be given as input to the players, and $W$ a \emph{set of allowed
  answers}.  Formally, $\sigma$ is an $m$-tuple $(G_1, \ldots, G_m)$
(where $m \in \mathbb{N}$ is the number of groups) of disjoint sets
$G_k \subseteq \bN{n}$ such that $\cup_k G_k = \bN{n}$, $q$ is an
$m$-tuple of bitstrings, and $W$ is a set of $m$-tuples of bitstrings.

The \emph{players} $P_1, \ldots, P_n$ are arbitrary (possibly
probabilistic) information-processing systems having an internal
state.  On each new input, a player generates an output depending on
this input (and possibly all previous inputs) and his internal state.

We will distinguish between two different settings: In the
\emph{classical} setting, the players are purely classical systems. In
this case, the initial values of their internal states $R_1, \ldots,
R_n$ (when the game starts) are given by a joint probability
distribution $P_{R_1 \cdots R_n}$ (in particular, the internal states
of the players might initially be correlated). An $n$-tuple of players
together with the probability distribution $P_{R_1 \cdots R_n}$ is
called a \emph{classical strategy} $\taucl$.

In the \emph{quantum} setting, the internal state of a player~$P_i$
additionally contains quantum information specified by the state of a
quantum system~$\cH_i$.  The player's inputs and outputs are still
classical, whereby the latter might depend on the (classical) outcomes
of measurements performed on $\cH_i$.  Before the start of the game,
the quantum systems $\cH_i$ are initialized with a quantum state $\kgs
\in \cH_1 \otimes \cdots \otimes \cH_n$ (the players' initial states
might thus be entangled). An $n$-tuple of quantum players together
with an $n$-partite initial state $\kgs$ (and, possibly an additional
classical probability distribution determining the initial values of
the classical parts of the players' internal states) defines a
\emph{quantum strategy $\tauqm$ (based on the state $\kgs$)}.

Let us now describe the rules of a game~$\cG_n$: First, an
\emph{instance} $(\sigma, q, W)$ is sampled according to the
probability distribution specified by $\cG_n$. The players are then
subdivided into groups defined by $\sigma$, i.e., a player $P_i$ is
said to \emph{belong to the group} $G_k$ if $i \in G_k$. Let us assume
that there are $m$ such groups.

The game consists of steps, where, in each step, each player takes
some input (which is identical for all players belonging to the same
group) and generates an output. In the first step, the players' inputs
are specified by the query $q = (q_1, \ldots, q_m)$ (where the
bitstring $q_k$ is given to all players in the group $G_k$).  Then,
the players communicate classically by generating outputs (in step
$t$) which are then (in the next step $t+1$) given as input to certain
other players.  In our model, a player can (in each step) choose
between two possibilities: his output is either sent to the players
within his group or it is broadcasted to all $n$
players.\footnote{Since the players are collaborating (and thus,
  privacy is no issue), this includes any type of communication among
  the players.  For instance, to send a certain message to \emph{one}
  specific player, the sender simply includes the address of the
  receiver (and possibly his own address) into the broadcasted
  message. Nevertheless, a distinction between broadcasted messages
  and messages sent to players within the group is needed for the
  definition of group broadcast complexity.}


The game runs until, after a certain number of steps, all $n$~players
are in a so-called halting state, indicated by a special output, where
additionally, at most one player in each group $G_k$ specifies a
\emph{final output string} $a_k$. (If there is no such player, we set
$a_k = \epsilon$ where $\epsilon$ is the empty string.) The game is
won if $(a_1, \ldots, a_m) \in W$.

As an example, consider the $n$-player game $\csG_n$ from
Section~\ref{sec:simpgame} (where we first omit the restriction that
the hint sent by the remaining players is limited to one bit). This
game obviously fits into the framework presented here: The two chosen
players ($P_i$ and $P_j$, for two different indices $i$ and $j$) each
form a one-player group while the remaining players are collected in a
third group. The partitioning of the $n$~players is thus determined by
the triple $\sigma_{ij} = ( \{i\}, \{j\}, \bN{n} \backslash \{i,j\}
)$.

The query $q$ specifies the information to be given to the players
when they are separated into groups. In our game, each player merely
learns whether he is among the chosen or the remaining ones. This can
be indicated by a bit, e.g., $q_s = (0, 0, 1)$. The game is won if the
outputs of the chosen players are different, i.e., the set of allowed
answers is $W_s = \{ (0,1,\epsilon), (1,0,\epsilon)\}$.

The choice of the two players~$P_i$ and~$P_j$ is random while the
query $q=q_s$ and the set of allowed answers $W=W_s$ is always the
same.  The game~$\csG_n$ is thus defined as the uniform distribution
over all triples $(\sigma_{ij}, q_s, W_s)$ where $1 \leq i < j \leq
n$.

\subsection{Broadcast Complexity} \label{sec:bc}



Let $\cG_n$ be a game and $P_1, \ldots, P_n$ a set of players.  Let
the information being broadcasted by player~$P_i$ in step $t$ be a
bitstring $b_{t,i}$ (where $b_{t,i} = \epsilon$ if $P_i$ does not
broadcast anything in this step).  In the next step $t+1$, this
information is given as input to all players in the form of a string
$\bb_{t} = b_{t,1} \| \cdots \|b_{t,n}$ being the concatenation of the
strings $b_{t,i}$ broadcasted in step~$t$.  These bitstrings $\bb_{t}$
must fulfill the requirement that any player, reading $\bb_{t}$
bitwise, is able to detect when the string terminates.\footnote{The
  length of any broadcasted string $\bb_{t}$ must thus either be fixed
  or be encoded into the string itself.} This technical point is
important when quantifying the amount of broadcasted bits since it
prevents information from being encoded into the length of
$\bb_{t}$.

The \emph{worst case group broadcast complexity} (or \emph{group
  broadcast complexity} for short) \emph{for a strategy $\tau$} is
defined as the maximum total number of bits broadcasted during the
game,
\[
  B(\cG_n,\tau) := \max \sum_t |\bb_{t}|\ ,
\]
where the maximum is taken over the whole randomness of the players
and their initial states (for probabilistic strategies) as well as the
randomness in the choice of the instance $(\sigma, q, W)$ of the
game~$\cG_n$.\footnote{Note that the worst case group broadcast
  complexity $B(\cG_n, \tau)$ only depends on the set of triples
  $(\sigma, q, W)$ having positive probability, but otherwise is
  independent of their exact distribution.}

The group broadcast complexity to win an $n$-player game $\cG_n$ is
generally smaller for quantum strategies than for classical
strategies. This motivates the following definition.

\begin{definition}
  The \emph{classical group broadcast complexity of a game $\cG_n$ for
    $n$ players}, $\Bcl(\cG_n)$, is the minimum value of
  $B(\cG_n,\taucl)$ where the minimum is taken over all classical
  strategies $\taucl$ to win $\cG_n$ with certainty.
  
  The \emph{quantum group broadcast complexity of $\cG_n$ with respect
    to an $n$-partite quantum state $\kgs$}, $\Bqm_{\kgs}(\cG_n)$, is
  defined similarly, but the minimum is taken over all winning quantum
  strategies $\tauqm$ based on $\kgs$.
\end{definition}

Note that, in the special case where for all instances of the game the
partitioning $\sigma$ is the trivial partitioning $(\{1\}, \ldots,
\{n\})$ (consisting of $n$ singleton sets), the group broadcast
complexity corresponds to Yao's definition of communication
complexity.

\subsection{A Separation Result} \label{sec:firstsep}

Let us again consider the example game~$\csG_n$ from
Section~\ref{sec:simpgame}.  It cannot be won classically if the
information a chosen player gets from the other players is restricted
to one bit (see Section~\ref{sec:simpcl}). On the other hand, the
quantum strategy presented in Section~\ref{sec:simpqm}, which is based
on a GHZ state, allows to always win the game with one hint bit.  This
is summarized by the following lemma, which additionally gives a lower
bound for the classical broadcast complexity.

\begin{lemma} \label{lem:simpcomp}
  The classical and the quantum group broadcast complexity of $\csG_n$
  satisfy
  \[
    \Bcl(\csG_n) \geq \log_2 \log_2 n 
    \quad \text{and} \quad
    \Bqm_{\kbs}(\csG_n) \leq 1\ ,
  \]
  respectively, where $\kbs$ is an $n$-partite GHZ state.
\end{lemma}

\begin{proof}
  The only missing part is the proof of the lower bound on the
  classical broadcast complexity $\Bcl(\csG_n)$, i.e., it has to be
  shown that for any classical strategy $\taucl$
  \[
    B(\csG_n, \taucl) \geq \log_2 \log_2 n\ .
  \]
  Since the group broadcast complexity $B(\csG_n, \taucl)$ is defined
  as a maximum taken over the randomness of the players, it suffices
  to prove this inequality to hold for any deterministic strategy
  (where each player's output is completely determined by his input).
  
  Let $\taucl$ be a deterministic strategy for winning $\csG_n$ with
  certainty. For any two indices $i, j$ ($1 \leq i < j \leq n$), let
  $m_{ij}$ be the concatenation of all strings $\bb_t$ (for
  $t=1,2,\ldots$) broadcasted by all players during the game if the
  instance $\sigma_{ij} = ( \{i\}, \{j\}, \bN{n} \backslash \{i,j\} )$
  has been chosen. Since each of the chosen players $P_i$ and $P_j$
  forms a one-player group, there is no communication within these
  groups. The only inputs of a chosen player are thus (in the first
  step) the query, which is always the bit $0$, and (in the subsequent
  steps) the broadcasted messages specified by $m_{ij}$.
  
  Let $M$ be the set containing the strings $m_{ij}$ (for any possible
  instance of the game~$\csG_n$), i.e.,
  \[
    M := \{m_{ij} : \; 1 \leq i < j \leq n \}\ .
  \]
  Clearly, the set $M$ contains (at least) one string $m_{ij}$ of
  length at least $\log_2 |M|$. Since the (maximum) length of the
  strings $m_{ij}$ is a lower bound for $B(\csG_n, \taucl)$, it
  suffices to prove that
  \begin{equation} \label{eq:comboundsimp}
    \log_2 |M| \geq \log_2 \log_2 n\ .
  \end{equation}
  
  The final output bit of any (deterministic) player $P_i$, when he is
  chosen, is fully determined by $m \in M$ (defining the sequence of
  broadcasted messages). Let $m_1, \ldots, m_l$ be the $l := |M|$
  elements of $M$. Furthermore, for each $i=1, \ldots, n$, let
  $\borg{i}$ be a bitstring of length $l$ where the $r$th bit
  $\borg{i}_r$ (for $r=1, \ldots, l$) is the output bit (or an
  arbitrary bit, if there is no such output) of $P_i$ given that the
  sequence of broadcasted messages is $m_r$.
  
  By assumption, the players always win the game. Consequently, for
  any instance $\sigma_{ij} = ( \{i\}, \{j\}, \bN{n} \backslash
  \{i,j\} )$, there must be a sequence of broadcasted messages $m \in
  M$ such that the output bits of the chosen players~$P_i$ and~$P_j$
  are different.  This is equivalent to say that the bitstrings
  $\borg{1}, \ldots, \borg{n}$ must all be different. Their length $l=
  |M|$ is thus lower bounded by $\log_2 n$, from which
  inequality~(\ref{eq:comboundsimp}) immediately follows.
\end{proof}

\subsection{A Generalized Pseudo-Telepathy Game and a Stronger Separation Result} \label{sec:gengame}


We will now consider an $n$-player game $\cbG_n$ for which the gap
between the classical and the quantum broadcast complexity is even
larger.

For any subset $C \subseteq \bN{n}$, $C=\{c_1, \ldots, c_k\}$ (where
$k = |C|$), let
\begin{eqnarray*}
    \sigma_C 
  & = & 
    (\{c_1\}, \ldots,\{c_k\}, \bN{n} \backslash C\})
  \\
    q_k 
  & = & 
    (\underbrace{0, \ldots, 0}_{k \text{ times}}, 1)
  \\
    W_k 
  & = & 
    \{(\underbrace{b_1, \ldots, b_k}_{k \text{ times}}, \epsilon) 
        : \; b_i \in \{0,1\}; \; 
        \sum_{i=1}^k b_i \equiv 1 \pmod{2} \} \ .
\end{eqnarray*}
The $n$-player game $\cbG_n$ (for $n \in \mathbb{N}$) is defined as
the uniform distribution over all triples $(\sigma_C, q_{|C|},
W_{|C|})$ with $C \subseteq \bN{n}$ and $|C| \equiv 2 \pmod{4}$.

Note that this game is very similar to the game~$\csG_n$ from
Section~\ref{sec:simpgame}: First, $k = 4 t + 2$ players (for some
random integer $t$) are randomly chosen. (The query bit is used to
indicate whether a player belongs to the chosen or the remaining
ones.) Each of the chosen players must then generate an output bit
$b_i$ such that the parity of all these bits is odd (in particular,
for $k=2$, the two bits must be different).

The $k$ chosen players each form a one-player group while another
group consists of the $n-k$ remaining players. A hint string sent by
the remaining players to the chosen players thus counts as inter-group
communication.  We will show that for players sharing a GHZ state one
single hint bit always suffices to win $\cbG_n$ while the classical
group broadcast complexity for this game is at least $\frac{1}{2}
\log_2 n - 2$.

\subsubsection{Classical Analysis}

The game~$\cbG_n$ can always be won classically with $\lceil \log_2 n
\rceil$ bits of inter-group communication. To see this, consider the
following strategy: A unique labeling bitstring $m_i$ of length
$\lceil \log_2 n \rceil$ is assigned to each player $P_i$ (for $i=1,
\ldots, n$).  During the game, the remaining players first communicate
within their group in order to find out the label $m_s$ of an
arbitrary player $P_s$ not belonging to their group ($m_s$ is thus the
label of a chosen player), and then broadcast $m_s$. Each of the
chosen players $P_i$ compares this message $m_s$ with his label $m_i$
and then generates a final output bit $b_i$ such that $b_i=1$ if (and
only if) $m_i=m_s$.\footnote{If all $n$ players have been chosen,
  there are no remaining players sending a bitstring to the chosen
  players.  To overcome this problem, one might think of a standard
  behavior for empty groups defined by the strategy.}  Then,
obviously, only player~$P_s$ outputs $1$, i.e., the game is won.

The classical broadcast complexity of $\cbG_n$ is thus at most $\lceil
\log_2 n \rceil$ bits. It turns out that, with any classical strategy
using less than (roughly) one half of this amount of inter-group
communication, there is a nonzero probability to lose the game.

\begin{lemma} \label{lem:gengamecl}
  In order to win the game~$\cbG_n$ classically with certainty, at
  least $\frac{1}{2} \log_2 n - 2$ bits of information have to be
  exchanged between the groups, i.e., $\Bcl(\cbG_n) \geq \frac{1}{2}
  \log_2 n - 2$.
\end{lemma}

\begin{proof}  
  The proof is analog to the proof of Lemma~\ref{lem:simpcomp}. The
  instances of the game~$\cbG_n$ are parameterized by subsets $C
  \subseteq \bN{n}$ with $|C| \equiv 2 \pmod{4}$.  For some fixed
  deterministic classical strategy, let
  \[
    M := \{m_C : C \subseteq \bN{n}; \; |C| \equiv 2 \pmod{4} \}\ ,
  \]
  where $m_C$ is the concatenation of all strings $\bb_t$ (for
  $t=1,2,\ldots$) broadcasted by the players given that the instance
  $(\sigma_C, q_{|C|}, W_{|C|})$ has been chosen.  Since $\log_2 |M|$
  is a lower bound for the group broadcast complexity (see proof of
  Lemma~\ref{lem:simpcomp}), it remains to be proven that
  \begin{equation} \label{eq:comboundgen}
    \log_2 |M| \geq \frac{1}{2} \log_2 n -2\ .
  \end{equation}
  
  Let $m_1, \ldots, m_l$ be the $l := |M|$ elements of $M$.
  Furthermore, define the $l$-bit strings $\borg{i}$ (for $i=1,
  \ldots, n$) as in the proof of Lemma~\ref{lem:simpcomp}: The $r$th
  bit $\borg{i}_r$ (for $r=1, \ldots, l$) is the output bit of player
  $P_i$ given that the sequence of broadcasted messages is $m_r$.  If
  the players win the game~$\cbG_n$ with certainty, then, for any
  allowed set $C$ defining an instance $(\sigma_C, q_{|C|}, W_{|C|})$,
  there must be a sequence of broadcasted messages $m \in M$ such that
  the parity of the output bits of the chosen players $P_i$ ($i \in
  C$) is odd.
  
  This requirement can again be formulated as a condition on the
  bitstrings $\borg{1}, \ldots, \borg{n}$: For all sets $C \subseteq
  \bN{n}$ with $|C| \equiv 2 \pmod{4}$ there exists an element $r \in
  \bN{l}$ such that $\sum_{i \in C} \borg{i}_r \equiv 1 \pmod{2}$.
  Lemma~\ref{lem:indep} (see appendix) states that the length $l =
  |M|$ of these bitstrings, which can be considered as elements of an
  $l$-dimensional vector space over $\GF(2)$, is lower bounded by
  $\sqrt{n}-2$. Since $l\geq 1$, this implies
  inequality~(\ref{eq:comboundgen}) and thus concludes the proof.
\end{proof}

\subsubsection{Quantum Analysis}

There is a quantum strategy to win $\cbG_n$ which exactly corresponds
to the winning strategy for the game from Section~\ref{sec:simpgame}.
However, the game~$\cbG_n$ allows for more possibilities on how
players might be chosen.  Therefore, for the proof of the following
lemma, a more general analysis than the one given in
Subsection~\ref{sec:simpqm} is needed.

\begin{lemma}
  To win the game~$\cbG_n$ with certainty using a GHZ state~$\kbs$,
  only one (classical) bit has to be exchanged between the groups,
  i.e., $\Bqm_{\kbs}(\cbG_n) \leq 1$.
\end{lemma}

\begin{proof}  
  Let $P_1, \ldots, P_n$ be $n$~players, each of them controlling a
  two-dimensional quantum system $\cH_1, \ldots, \cH_n$, respectively.
  Furthermore, let $\{\ve{0}, \ve{1}\}$ be an orthonormal basis of
  $\cH_i$ (for all $i=1, \ldots, n$) and define the diagonal basis
  $\{\vf{0}, \vf{1}\}$ and the circular basis $\{\vg{0}, \vg{1}\}$ as
  in Subsection~\ref{sec:simpqm}. The GHZ state $\kbs$ initially
  shared by the players is then given by~(\ref{eq:ghzdef}).
  
  The strategy of the players is as follows: If a player is among the
  remaining ones (i.e., if he gets a bit $1$ as query input in the
  first step), he measures his quantum system $\cH_i$ with respect to
  the diagonal basis $\{\vf{0}, \vf{1}\}$ and sends the result of this
  measurement to the other players within his group (i.e., to the
  other remaining players). One of the remaining players then
  broadcasts a bit $b$ depending on whether an even ($b=0$) or an odd
  ($b=1$) number of them got the measurement outcome $\vf{1}$.
  
  If a player~$P_i$ is among the chosen players (i.e., his first input
  is~$0$), he reads the bit $b$ broadcasted by the group of remaining
  players and then measures his system $\cH_i$, depending on this bit,
  using either the diagonal basis $\{\vf{0}, \vf{1}\}$ (if $b=1$) or
  the circular basis $\{\vg{0}, \vg{1}\}$ (if $b=0$). He then simply
  outputs a bit indicating the outcome of this measurement.
  
  The broadcast complexity of this strategy is obviously $1$. It thus
  remains to be verified that, for all instances $(\sigma, q, W)$ of
  $\cbG_n$, the players $P_1, \ldots, P_n$ win with certainty.  By the
  symmetry of the game and the described strategy, the analysis is
  exactly the same for all instances. We can thus, without loss of
  generality, restrict to one instance (for each possible $k$), namely
  $(\sigma_{C}, q_{k}, W_{k})$ where $C = \{1, \ldots, k \}$, i.e.,
  $P_1, \ldots, P_k$ are the chosen players while $P_{k+1}, \ldots,
  P_n$ are the remaining ones.
  
  It is easy to check that the vectors
  \begin{equation}
    \ket{v_{b_1 \cdots b_n}} := 
    \begin{cases}
        \vf{b_1} \otimes \cdots \otimes \vf{b_k}
        \otimes \vf{b_{k+1}} \otimes \cdots \otimes \vf{b_n} 
      & \text{if } \sum_{i=k+1}^n b_{i} \equiv 1 \pmod{2} \\
        \vg{b_1} \otimes \cdots \otimes \vg{b_k}
        \otimes \vf{b_{k+1}} \otimes \cdots \otimes \vf{b_n} 
      & \text{if } \sum_{i=k+1}^n b_{i} \equiv 0 \pmod{2}
    \end{cases}
  \end{equation}
  (for all $(b_1, \ldots, b_n) \in \{0,1\}^n$) build an orthonormal
  basis of $\cH_1 \otimes \cdots \otimes \cH_n$.  Note that these
  vectors are the products of the measurement bases used by the
  players when following the described strategy, where $b_{k+1},
  \ldots, b_n$ are the measurement outcomes of the remaining players
  and $b_1, \ldots, b_k$ are the final output bits of the chosen
  players.  The probability that the chosen players $P_1, \ldots, P_k$
  have output $b_1, \ldots, b_k$, respectively, is thus given by
  \[  
      p_{b_1 \cdots b_k} 
    := 
      \sum_{(b_{k+1}, \ldots, b_n) \in \{0,1\}^{n-k}} p_{b_1 \cdots b_n}
  \]
  where 
  \[
    p_{b_1 \cdots b_n} := |\spr{\bs}{v_{b_1 \cdots b_n}}|^2\ .
  \] 
  It remains to be shown that the probability for the output of the
  chosen players not being contained in $W_{k}$ is zero, i.e.,
  \begin{equation} \label{eq:impl}
    \sum_{i=1}^k b_i \equiv 0 \pmod{2}
    \quad \Longrightarrow \quad
    p_{b_1 \cdots b_n} = 0\ .
  \end{equation} 
  
  Let us first assume that $\sum_{i=k+1}^n b_{i} \equiv 1 \pmod{2}$.
  We then have
  \[
  \begin{split}
      \spr{\bs}{v_{b_1 \cdots b_n}}
     = 
      2^{-\frac{n+1}{2}}
        &\bigl(\veb{0} \otimes \cdots \otimes \veb{0} 
         + \veb{1} \otimes \cdots \otimes \veb{1}\bigr) 
    \\ & 
      \cdot
        \bigl(\ve{0} + (-1)^{b_1} \ve{1}) \otimes \cdots 
         \otimes (\ve{0} + (-1)^{b_n} \ve{1} \bigr)
  \end{split}
  \]
  and thus
  \[
    p_{b_1 \cdots b_n} = 2^{-(n+1)} | 1 + (-1)^{\sum_{i=1}^n b_i} |^2\ .
  \]
  From the assumption on $b_{k+1}, \ldots, b_n$ it follows immediately
  that for $b_1, \ldots, b_k$ satisfying the left side of
  implication~(\ref{eq:impl}), the sum in the exponent becomes odd,
  i.e., the probability $p_{b_1 \cdots b_n}$ is zero.
  
  Assume now that $\sum_{i=k+1}^n b_{i} \equiv 0$. Then
  \[
  \begin{split}
      \spr{\bs}{v_{b_1 \cdots b_n}}
    = 
      2^{-\frac{n+1}{2}} 
        & (\veb{0} \otimes \cdots \otimes \veb{0} 
         + \veb{1} \otimes \cdots \otimes \veb{1}) 
\\  &   \cdot
        (\ve{0} + (-1)^{b_1} i \, \ve{1}) \otimes \cdots 
         \otimes (\ve{0} + (-1)^{b_k} i \, \ve{1}) 
\\  & \qquad \qquad
        \otimes
        (\ve{0} + (-1)^{b_{k+1}} \ve{1}) \otimes \cdots 
         \otimes (\ve{0} + (-1)^{b_n} \ve{1})
  \end{split}
  \]
  and hence
  \[
      p_{b_1 \cdots b_n} 
    = 2^{-(n+1)} | 1 + i^k (-1)^{\sum_{i=1}^n b_i} |^2
    = 2^{-(n+1)} | 1 - (-1)^{\sum_{i=1}^n b_i} |^2\ ,
  \]
  where the second equality follows from $k \equiv 2 \pmod{4}$.  From
  the assumption on $b_{k+1}, \ldots, b_n$ we can again conclude that
  implication~(\ref{eq:impl}) is satisfied.
\end{proof}

\section{Conclusion}


The classical outcomes of measurements performed on an entangled
quantum state can generally not be explained by local classical
randomness. This non-local property of quantum mechanics is
demonstrated by the pseudo-telepathy game proposed in
Section~\ref{sec:simpgame}: A simple task, which \emph{obviously}
cannot be accomplished by separated classical players, is solvable by
players sharing quantum entanglement.

The non-locality of an $n$-partite quantum state is often
characterized by the amount of communication needed by $n$ separated
classical systems for simulating the outcomes of local measurements
performed on the respective parts of the state.  There are clearly
several ways to quantify this communication, each revealing a
different aspect of the non-local information contained in the state.
Contrary to the approach of Cleve and Buhrman~\cite{CB97}, we consider
the information exchanged between certain groups of systems instead of
counting the overall communication. This leads to an alternative
quantification of non-local information and, consequently, to new
separation results.  They cannot directly be compared with the results
based on Yao's model (as for instance~\cite{BCD01}), but rather unveil
another facet of the nature of entanglement as well as the gap between
quantum and classical correlation.

The results obtained in Section~\ref{sec:quant} are formulated in
terms of communication complexity with respect to certain games. The
difference between the amount of classical communication needed for
classical and quantum players, respectively, to win such games
directly lead to lower bounds for the communication needed to simulate
quantum states.  For instance, the separation stated by
Lemma~\ref{lem:simpcomp} implies that for the classical simulation of
two separated two-dimensional quantum systems sharing a GHZ state with
$n-2$ other systems, at least $\Omega(\log \log n)$ bits of additional
information are necessary.

It is one of the goals of this paper to shed some light on the nature
of quantum entanglement, a phenomenon which is not yet completely
understood. While separation results, as the ones presented here, can
be seen as lower bounds for the amount of non-local information
contained in entangled quantum states, some work has been done to
determine the maximal communication being necessary for an exact
simulation of such states by classical systems (see, e.g.,
\cite{BCT99}).  It is, however, still an open problem to find the most
accurate way to characterize entanglement between quantum systems in
terms of classical communication.

\section{Acknowledgments}

The authors thank Gilles Brassard, Nicolas Gisin, Ueli Maurer, and
Alain Tapp for interesting discussions. The first author was partially
supported by the Swiss National Science Foundation (SNF), and the
second author was partially supported by the Natural Sciences and
Engineering Research Council of Canada (NSERC).

\newpage

\appendix

\section*{Appendix}

\begin{lemma} \label{lem:indep}
  Let $\borg{1}, \ldots, \borg{n}$ be $n$ vectors of an
  $l$-dimensional vector space over $\GF(2)$.  If for all $C \subseteq
  \bN{n}$ with $|C| \equiv 2 \pmod 4$
\vspace{-2mm}
  \begin{equation} 
    \sum_{i \in C} \borg{i} \neq \mathbf{0}
  \end{equation}
  then
\vspace{-2mm}
  \begin{equation} \label{eq:lbnd}
    l \geq \sqrt{n}-2\ .
  \end{equation}
\end{lemma}

\begin{proof}  
  Each element of a $d$-dimensional vector space over $\GF(2)$ can
  naturally be identified with a bitstring of length $d$, and vice
  versa. In the following, we will thus alternately speak of vectors
  and bitstrings, always meaning the same object.
  
  The idea is to append additional bits to the bitstrings $\borg{i}$
  (for $i = 1, \ldots, n$) in order to obtain longer bitstrings
  $\bone{i}$ and $\bfin{i}$. These bits are chosen in such a way that
  the resulting bitstrings $\bone{i}$ or $\bfin{i}$, considered as
  vectors, are linearly independent. This will lead to a lower bound
  on their length which finally allows to derive a lower bound on $l$.
  
  Defining (for all $i = 1, \ldots, n$)
  \[
    \bone{i} := 1 \| \borg{i}
  \]
  (where $\|$ is the concatenation of strings) we have, for any set $I
  \subseteq \bN{n}$,
  \begin{equation} \label{eq:condbone}
      \sum_{i \in I} \bone{i} = \mathbf{0} 
    \quad \Longrightarrow \quad
      |I| \equiv 0 \pmod 4\ .
  \end{equation}
  This can be seen as follows: If $|I|$ is odd, the bits in the first
  position of the strings $\bone{i}$ (which are all equal to $1$) will
  sum up to $1$. On the other hand, if $|I| \equiv 2 \pmod 4$, then,
  by the assumption of the lemma, the sum $\sum_{i \in I} \borg{i}$ is
  nonzero.
  
  For a given family $\mathcal{A}$ of disjoint nonempty subsets of
  $\bN{n}$, let $\bar{A}$ be the set of elements not contained in any
  of these subsets, $\bar{A} := \bN{n} \backslash \cup_{A \in
    \mathcal{A}} A$, and set $\bar{\mathcal{A}} := \mathcal{A} \cup
  \{\bar{A}\}$. It is easy to see that there exists such a family
  $\mathcal{A}$ satisfying the following condition: For any nonempty
  set $B$ with $B \subseteq A$ for some $A \in \bar{\mathcal{A}}$
  \begin{equation} \label{eq:condB}
      \sum_{i \in B} \bone{i} = \mathbf{0} 
    \quad \Longleftrightarrow \quad
      B \in \mathcal{A}\ ,
  \end{equation}
  i.e., the sets $A \in \mathcal{A}$ are minimal sets of indices such
  that the vectors $\bone{i}$ for $i \in A$ are linearly dependent.
  Note that, from~(\ref{eq:condbone}), we have
  \begin{equation} \label{eq:condA}
    |A| \equiv 0 \pmod 4
  \end{equation}
  for all $A \in \mathcal{A}$.
  
  Let $\mathcal{A}$ be a family of sets satisfying
  condition~(\ref{eq:condB}). We will distinguish two cases.
  
  First, assume that there is a set $A \in \bar{\mathcal{A}}$ such
  that $|A| > \sqrt{n}$. Let $V$ be an arbitrary subset of $A$ with
  $|V| = |A|-1$. It follows directly from condition~(\ref{eq:condB})
  that the vectors $\bone{i}$ for $i \in V$ form a set of $|A|-1$
  linearly independent vectors. Consequently, their length $l+1$ must
  satisfy $l+1 \geq |A|-1$ which immediately implies~(\ref{eq:lbnd}).
  
  Assume now that $A \leq \sqrt{n}$ for all $A \in \bar{\mathcal{A}}$.
  It follows directly that $|\mathcal{A}| \geq \sqrt{n}-1$.  Let, for
  $i = 1, \ldots, n$,
  \[
    \bmax{i} := e_i \| \bone{i} 
  \]
  where $e_i$ is the bitstring of length $n$ which has a bit $1$ at
  the $i$th position and zeros at all other positions. Furthermore,
  for all $A \in \mathcal{A}$, let $r_A$ be an arbitrary element of
  $A$. Define $\bfin{i}$ (for $i=1, \ldots, n$) as the bitstring which
  is identical to $\bmax{i}$ except that the bits at positions $r_A$
  (for all $A \in \mathcal{A}$) are omitted. Since the strings
  $\bmax{i}$ have length $n+1+l$, the strings $\bfin{i}$ obviously
  have length $l' = n+1+l-|\mathcal{A}| \leq n-\sqrt{n}+l+2$.
  
  If the strings $\bfin{i}$ (for $i = 1, \ldots, n$) form a set of $n$
  linearly independent vectors, then $l' \geq n$, i.e.,
  \[
    n-\sqrt{n}+l+2 \geq n
  \]
  which again implies~(\ref{eq:lbnd}).
  
  It thus remains to be shown that the bitstrings $\bfin{1}, \ldots,
  \bfin{n}$ are indeed linearly independent.  Assume by contradiction
  that there is a nonempty set $I \subseteq \bN{n}$ such that
  \begin{equation} \label{eq:sumzero}
    \sum_{i \in I} \bfin{i} = \mathbf{0}\ . 
  \end{equation}
  To show that this leads to a contradiction, we will distinguish
  three cases.
  \begin{enumerate}[(a)]
    
  \item $|I| \not\equiv 0 \pmod 4$: From condition~(\ref{eq:condbone})
    the sum $\sum_{i \in I} \bone{i}$ is nonzero. Since the last $n+1$
    bits of $\bfin{i}$ correspond to $\bone{i}$, this obviously
    contradicts equation~(\ref{eq:sumzero}).
    
  \item $|I| \equiv 0 \pmod 4$ and $I \cap \bar{A} \neq \emptyset$:
    Let $r$ be an element of the intersection $I \cap \bar{A}$.  By
    definition, there is exactly one bitstring $\bmax{i}$ with $i \in
    I$ having a bit $1$ at the $r$th position, namely $\bmax{r}$.
    Note that the bit of $\bmax{i}$ at position $r$ corresponds to a
    bit of $\bfin{i}$ at some position $r'$ (in the construction of
    $\bfin{i}$ from $\bmax{i}$ only bits with an index in sets $A$
    with $A \in \mathcal{A}$ are omitted).  Consequently, the $r'$th
    bit of the sum in~(\ref{eq:sumzero}) is $1$.
    
  \item $|I| \equiv 0 \pmod 4$ and $I \cap \bar{A} = \emptyset$: Since
    $I$ is nonempty, there exists a set $A \in \mathcal{A}$ such that
    $I \cap A \neq \emptyset$. Assume that $|I \cap A| > 1$, i.e.
    there are at least two different indices $r_1$ and $r_2$ in $I
    \cap A$. Consequently, the sum $\sum_{i \in I} \bmax{i}$ has a bit
    $1$ at position $r_1$ and $r_2$. By the construction of the
    strings $\bfin{i}$, at least one of these bits corresponds to a
    bit in the sum in~(\ref{eq:sumzero}) which can thus not be zero.
  
    It remains to be shown that $|I \cap A| > 1$. Using the fact that
    $\bone{i}+\bone{i} = \mathbf{0}$ (over $\GF(2)$) we have
    \[
        \sum_{i \in A/I \cup I/A} \bone{i} 
      = \sum_{i \in A} \bone{i} + \sum_{i \in I} \bone{i} = \mathbf{0}
    \]
    where the last equality follows from $A \in \mathcal{A}$ and
    condition~(\ref{eq:condB}) as well as from
    assumption~(\ref{eq:sumzero}). With condition~(\ref{eq:condbone})
    this implies that
    \begin{equation} \label{eq:AIno}
      |A/I \cup I/A| \equiv 0 \pmod 4\ .
    \end{equation}
    On the other hand, using $(\ref{eq:condA})$ and $|I| \equiv 0
    \pmod 4$,
    \[
      |A/I \cup I/A| = |A| + |I| - 2 |A \cap I|
      \equiv -2 |A \cap I| \pmod 4\ .
    \]
    Together with (\ref{eq:AIno}) we conclude that $|A \cap I|$ must
    be even and thus, since the set $A \cap I$ is nonempty, we have
    $|A \cap I|>1$.
  \end{enumerate}
\end{proof}

\bibliographystyle{hplain}
 
\bibliography{entgame}

\end{document}